\documentclass[11pt]{article}
\pdfoutput=1

\usepackage{jcappub}

\usepackage{amsfonts} 
\usepackage{amsmath} 

\usepackage{graphicx} 
\usepackage{float}
\usepackage[margin=1in]{geometry}
\usepackage{float}
\usepackage{tabularx}
\usepackage{array}
\usepackage{epstopdf}

\usepackage{caption}
\usepackage{subfig}

\title{Random Functions via Dyson Brownian Motion: Progress and Problems}

\author{Gaoyuan Wang,}
\author{Thorsten Battefeld}
\affiliation{Institute for Astrophysics, University of Goettingen, Friedrich Hund Platz 1, D-37077 Goettingen, Germany}

\emailAdd{gaoyuan.wang@stud.uni-goettingen.de,\\
tbattefe@gmail.com}

\abstract{ We develope a computationally efficient extension of the Dyson Brownian Motion (DBM) algorithm to generate random function in $C^2$ locally. We further explain that random functions generated via DBM show an unstable growth as the traversed distance increases. This feature restricts the use of such functions considerably if they are to be used to model globally defined ones. The latter is the case if one uses random functions to model landscapes in string theory. We provide a concrete example, based on a simple axionic potential often used in cosmology, to highlight this problem and also offer an ad hoc modification of DBM that suppresses this growth to some degree.
}

\arxivnumber{1607.02514}

\begin{document}
\maketitle
\flushbottom

\section{Introduction}
\label{sec:intro}

Generating random functions locally via Dyson Brownian Motion (DBM) was first introduced in \cite{Marsh:2013qca} following prior work by Dyson \cite{Dyson:62}, see \cite{Mehta:1991} for a textbook, and applied to potentials in cosmology in \cite{Marsh:2013qca,Battefeld:2014qoa,Dias:2016slx}. The algorithm starts with a Taylor expansion of the potential to second order and, after moving a set distance from the initial expansion point, adding random perturbations to the matrix of second derivatives\footnote{Dyson Brownian motion \cite{Dyson:62} was developed primarily to introduce a ``time'' dependence into Random Matrix Theory, which enables the computation of transition probabilities. Analytic methods to compute such probabilities were derived in \cite{Pedro:2016jyd}.}. Thus, a random function is generated locally along a trajectory. The algorithm as introduced in \cite{Marsh:2013qca} generates a function in $C^1$ along a trajectory\footnote{The continuum limit of the algorithm yields a function in $C^2$. If implemented numerically and the step length of the DBM function is kept in line with the discretization used elsewhere, the functions introduced in \cite{Marsh:2013qca} are indistinguishable from functions in $C^2$. However, using functions that are in $C^2$ by construction can be computationally advantageous.} with a Hessian in the Gaussian orthogonal ensemble. In \cite{Battefeld:2014qoa}, the method was generalized to generate functions in any differentiability class by relegating the perturbations to a higher order derivative tensor, with particular emphasis on functions in $C^2$. Such functions are of particular interest to model the potential in inflationary cosmology, see \cite{Baumann:2014nda,Marsh:2015xka} for recent reviews, while enabling numerical computation of the power-spectrum of cosmological fluctuations \footnote{In \cite{Dias:2016slx}, semi-analytic methods were used to compute observables. } without generating ringing or requiring a step size for DBM in line with the one used for the discretization of the required differential equations. However, the methods introduced in  \cite{Battefeld:2014qoa} are computationally more demanding than ordinary DBM, since a coordinate rotation is needed at each step.

In this brief technical note we improve upon the algorithm presented in \cite{Battefeld:2014qoa} to generate a function in $C^2$ without the need to rotate the Hessian, see Sec.~\ref{Sec:extDBM}. As a consequence, field spaces with more than 100 fields, as needed to model certain landscapes in string theory/cosmology, can be handled. We also explain a severe shortcoming of functions generated via DBM or its generalization: while the elements of the Hessian have a stable distribution, the function itself performs a random walk and is thus unbounded. Simply using the appropriate distribution for a Hessian in a bounded potential, such as the axionic one investigated in \cite{Wang:2015rel}, instead of one in the Gaussian orthogonal ensemble does not alleviate the problem. This shortcoming renders DBM potentials ill suited to model many landscapes of interest in cosmology if one is interested in going beyond the first coherent patch. We explain this shortcoming in a concrete case-study in Sec.~\ref{Sec:axionic} and provide a minor modification of DBM that can suppress the unstable growth to some degree. 

We refer the interested reader to \cite{Marsh:2013qca,Battefeld:2014qoa,Dias:2016slx} for a more pedagogical introduction to DBM and cosmological implications, while focusing on technical aspects in this note.

\section{Generating $C^2$ Random Potentials \label{sec:C2potentials}}

\subsection{Dyson Brownian Motion Potentials \label{DBMreview}}

To generate random functions, Dyson \cite{Dyson:62}, see also \cite{Marsh:2013qca}, proposed to consider the Hessian to be a stochastic variable whose change after a step
\begin{eqnarray}
\delta s \equiv \|\delta {\phi}^a\| \equiv \sqrt{\delta \phi_a\delta \phi^a}
\end{eqnarray}
along some path in a $D$-dimensional field space\footnote{  We use Einstein's summation convention over field indices and consider a flat field space metric. } with fields $\phi_a$, $a=1\dots D$ is given by
 \begin{eqnarray}
\delta \mathcal{H}_{ab}=\delta A_{ab} -\mathcal{H}_{ab}\frac{\delta s}{\Lambda_{\text{h}}}\,.
\end{eqnarray}
Here, $\Lambda_{\text{h}}$ is the horizontal correlation length of the Hessian, the $\delta A_{ab}$ are $D(D + 1)/2$ zero-mean stochastic variables and the term $\propto -\mathcal{H}_{ab}$ is  a restoring force. Due to the latter, the Hessian is a (symmetric) matrix in the Gaussian Orthogonal Ensamble (GOE, Wigner matrix).  
The first two moments of the Hessian obey \cite{Marsh:2013qca}
\begin{eqnarray}
 \langle \delta \mathcal{H}_{ab}|_{p_1}\rangle &=& -\mathcal{H}_{ab}|_{p_0} \frac{\delta s}{\Lambda_{\text{h}}}\,, \label{wignercondition1} \\
\langle (\delta \mathcal{H}_{ab})^2\rangle &=& (1+\delta_{ab}) \frac{\delta s}{\Lambda_{\text{h}}}\sigma^2\,, \label{wignercondition2} 
\end{eqnarray}
 where $\sigma$ is the standard deviation of the GOE.

To apply DBM and construct a random potential\footnote{Since we have cosmological applications in mind, we denote the random function with $V$ and refer to it as a potential. Similarly, we refer to the variables of said function as fields.}, consider the Taylor expansion at some initial point $p_0$
\begin{eqnarray}
V= \Lambda_{\text{v}}^4 \sqrt{D}\left[v_0 + v_a\tilde{\phi}^a+ \frac{1}{2}v_{ab}\tilde{\phi}^a\tilde{\phi}^b+\dots\right]\,, \label{taylorpotential}
\end{eqnarray}
where $\Lambda_{\text{v}}$ sets the vertical scale
and $\tilde{\phi}^a \equiv \phi^a /\Lambda_{\text{h}}$.  Both, $\Lambda_{\text{v}}$ and $\Lambda_{\text{h}}$, have units of mass.
Truncating the series at second order and evaluating the potential at a nearby point $p_1$ a distance $\|\delta \tilde{\phi}^a\| \ll 1$ from the initial point $p_0$ away, we get
\begin{eqnarray}
 v_{ab}\!\mid_{p_1} &=&v_{ab}\!\mid_{p_0}+\delta v_{ab}\!\mid_{p_0}\,,
\end{eqnarray}
 where $\delta v_{ab}\!\mid_{p_0}$ are elements of a random matrix with
\begin{eqnarray}
 \langle \delta v_{ab}\!\mid_{p_n}\rangle &=& -v_{ab}\!\mid_{p_{n-1}} \frac{\|\delta \phi^a\|}{\Lambda_{\text{h}}}\,,\label{condition1}\\
 \langle (\delta v_{ab}\!\mid_{p_n})^2\rangle &=& (1+\delta_{ab}) \frac{\|\delta \phi^a\|}{\Lambda_{\text{h}}}\sigma^2\,.\label{condition2}
 \end{eqnarray}
Repeated application along some path results in a locally defined, random potential $V\in C^1$. See \cite{Marsh:2013qca} for a more detailed discussion.

\subsection{Extended Dyson Brownian Motion Potentials, $V\in C^2$ \label{Sec:extDBM}}

In \cite{Battefeld:2014qoa}, the DBM algorithm was extended to generate random potentials $V \in C^k$ with $k=2,3,\dots$, by relegating perturbations to higher order derivative tensors. Two methods were introduced in \cite{Battefeld:2014qoa} for $V\in C^2$, both computationally demanding, since they require a rotation of the coordinate system after each step. Thus, an application to large $D$ is cumbersome. Here, we present an algorithm that doesn't require any coordinate transformations and is therefore efficient enough to generate random potentials with $D=\mathcal{O}(100)$.

Let us start by expanding the potential to third order,
\begin{eqnarray}
V=\Lambda_{\text{v}}^4\sqrt{D}\left(v_0+v_a\tilde{\phi}^a+\frac{1}{2}v_{ab}\tilde{\phi}^a\tilde{\phi}^b+\frac{1}{6}v_{abc}\tilde{\phi}^a\tilde{\phi}^b\tilde{\phi}^c\right)\,.
\end{eqnarray}
 The potential at a nearby point $p_1$ close to $p_0$ can thus be written as
\begin{eqnarray}
v_0|_{p_1}&=&v_0 |_{p_0}+v_a|_{p_0}\delta\tilde{\phi}^a+\frac{1}{2}v_{ab}|_{p_0}\delta\tilde{\phi}^a\delta\tilde{\phi}^b+\frac{1}{6}v_{abc}|_{p_0}\delta\tilde{\phi}^a\delta\tilde{\phi}^b\delta\tilde{\phi}^c+... \, ,\\
v_a |_{p_1}&=&v_a |_{p_0}+v_{ab}|_{p_0}\delta\tilde{\phi}^b+\frac{1}{2}v_{abc}|_{p_0}\delta\tilde{\phi}^b\delta\tilde{\phi}^c+... \, ,\\
v_{ab}|_{p_1}&=&v_{ab}|_{p_0}+v_{abc}|_{p_0}\delta\tilde{\phi}^c+... \, ,\\
v_{abc}|_{p_1}&=&v_{abc}|_{p_0}+\delta v_{abc}|_{p_0}\,.
\end{eqnarray}
Note that the random variable enters now in the tensor of third derivatives, $v_{abc}$. However, we still want the Hessian at well seperated points to act as a random variable with the same first two moments as in (\ref{condition1}) and (\ref{condition2}).
To this end, we define the $\delta v_{abc}$ as random variables as in \cite{Battefeld:2014qoa}, leading to
\begin{eqnarray}
\langle \delta v_{ab}|_{p_n}
\rangle&=&\langle v_{abc}|_{p_{n}}\delta\tilde{\phi}^c \rangle \,\\
&=&\langle (v_{abc}|_{p_{n-1}}+\delta v_{abc}|_{p_{n-1}})\delta\tilde{\phi}^c \rangle \,\\
&=& v_{abc}|_{p_{n-1}}\delta\tilde{\phi}^c+\langle \delta v_{abc}|_{p_{n-1}}\rangle \delta\tilde{\phi}^c  \,\\
&\equiv&-v_{ab}|_{p_{n-1}}\frac{\left\lVert\delta\phi^a\right\lVert}{\Lambda_{\text{h}}}\label{constrains1}\,,\\
\mathrm{Var}(\delta v_{ab}|_{p_n})&=&\mathrm{Var}( v_{abc}|_{p_{n}}\delta\tilde{\phi}^c )\,\\
&=&\mathrm{Var}( v_{abc}|_{p_{n-1}}\delta\tilde{\phi}^c )+\mathrm{Var}( \delta v_{abc}|_{p_{n-1}}\delta\tilde{\phi}^c )\,\\
&=&\mathrm{Var}( \delta v_{abc}|_{p_{n-1}}){\delta\tilde{\phi}^c}^2 \,\\ &\equiv&\frac{(1+\delta_{ab})\left\lVert\delta\phi^a\right\lVert\sigma^2}{\Lambda_{\text{h}}}-(v_{ab}|_{p_{n-1}}\frac{\left\lVert\delta\phi^a\right\lVert}{\Lambda_{\text{h}}})^2\,.
\label{constrains2}
\end{eqnarray}
Here, $v_{abc}$ and thus $\delta v_{abc}$ are symmetric under permutations of $a,b\, \text{and}\, c$. Note that the means and variances of $\delta v_{abc}$ appear in a sum, which lead to the use of rotations in \cite{Battefeld:2014qoa} to either align one coordinate axis with the preceding step vector, or to diagonalise the Hessian and extract conditions on the means and variances of $\delta v_{abc}$. Here, we make use of the fact that the above system of equations is under-determined to provide a more efficient algorithm.

For the $D$ dimensional case, $\delta v_{abc}$ is a $D \times D \times D$ totally symmetric tensor. We determine the mean and variance for each element separately, using the constraints given in (\ref{constrains1}) and (\ref{constrains2}) according to the following algorithm:

\begin{enumerate}
\item Set $a=1$ and determine the first $D \times D$ elements in $\delta v_{1bc}$. Set the off-diagonal entries of this matrix to zero and obtain a unique value for the diagonal entries using (\ref{constrains1}) and (\ref{constrains2}).
\item Consider $\delta v_{2bc}$. Elements of the row $\delta v_{21c}$ and column $\delta v_{2c1}$ inherit their values from $v_{1ab}$ because of the symmetry condition. The remaining $(D-1)\times(D-1)$ elements show up for the first time. We set all of the off-diagonal elements of this $(D-1)\times (D-1)$ matrix to zero and determine the mean and variance of the diagonal elements according to (\ref{constrains1}) and (\ref{constrains2}).
\item Proceed with  $\delta v_{ibc}, i \in [1,D]$ accordingly. Set rows $\delta v_{ijc}$ and columns $v_{icj}$, $j=1,...,i-1$ in line with its permutations according to the symmetry conditions. For the remaining $(D-i+1) \times (D-i+1)$ matrix, set the off-diagonal elements to zero and calculate the mean and variance of the diagonal elements using (\ref{constrains1}) and (\ref{constrains2}).
\item If the step length in a certain direction is too small, instabilities can arise due to numerical errors. 
Besides increasing the computational precision, and thus the computational cost, one may also include a lower limit for the path in each coordinate direction, $\delta\tilde{\phi}_{\text{limit}}$.  
If $\delta \tilde{\phi}^c<\delta\tilde{\phi}_{\text{limit}}$, one may simply set $\delta \tilde{\phi}^c\equiv \delta\tilde{\phi}_{\text{limit}}$ and avoid any computational problems.  A plausible choice is, for instance,
 $\delta\tilde{\phi}_{\text{limit}}=\delta s/(10\Lambda_{\text{h}} \sqrt{D})$. This lower limit will alter the trajectory slightly, which, depending on the question one wishes to investigate, can be acceptable.
\end{enumerate}

\subsection{Demonstration of the Algorithm \label{Sec:algorithm}}
Let us demonstrate the above procedure for $D=3$.

\paragraph{1.} We set all the off-diagonal elements of $\delta v_{1bc}$ to zero and choose the diagonal elements via 
\begin{eqnarray}
\langle \delta v_{111}|_{p_n}\rangle&=&\Big(-v_{11}|_{p_{n}}\frac{\left\lVert\delta\phi^a\right\lVert}{\Lambda_{\text{h}}}-v_{11c}|_{p_n}\delta\tilde{\phi}^c\Big){\delta\tilde{\phi}^1}^{-1} \,,\\
\mathrm{Var}(\delta v_{111}|_{p_n})&=&\Big(\frac{(1+\delta_{11})\left\lVert\delta\phi^a\right\lVert\sigma^2}{\Lambda_{\text{h}}}-(v_{11}|_{p_{n}}\frac{\left\lVert\delta\phi^a\right\lVert}{\Lambda_{\text{h}}})^2\Big){\delta\tilde{\phi}^1}^{-2} \,,\\
\delta v_{111}|_{p_n}&\in&\mathcal{N}\Big(\langle \delta v_{111}|_{p_n}\rangle,\mathrm{Var}(\delta v_{111}|_{p_n})\Big)\,.
\label{disexample}
\end{eqnarray}
The mean and variance for $\delta v_{122}$ and $\delta v_{133}$ are chosen analogously.

\paragraph{2.} For $\delta v_{2bc}$, we set the first row and the first column in line with their symmetric partners. We set the off-diagonal elements of the remaining $2 \times 2$ sub-matrix to zero and solve (\ref{constrains1}) as well as (\ref{constrains2}) for the diagonal ones
\begin{eqnarray}
\langle \delta v_{222}|_{p_n}\rangle&=&\Big(-v_{22}|_{p_{n}}\frac{\left\lVert\delta\phi^a\right\lVert}{\Lambda_{\text{h}}}-\langle \delta v_{221}|_{p_n}\rangle \delta\tilde{\phi}^1-v_{22c}|_{p_n}\delta\tilde{\phi}^c\Big){\delta\tilde{\phi}^2}^{-1} \,,\\
\mathrm{Var}(\delta v_{222}|_{p_n})&=&\Big(\frac{(1+\delta_{22})\left\lVert\delta\phi^a\right\lVert\sigma^2}{\Lambda_{\text{h}}}-(v_{22}| _{p_{n}}\frac{\left\lVert\delta\phi^a\right\lVert}{\Lambda_{\text{h}}})^2 \nonumber \\
&& -\mathrm{Var}(\delta v_{221}|_{p_n})(\delta\tilde{\phi}^1)^2\Big) {\delta\tilde{\phi}^2}^{-2} \,,\\
\delta v_{222}|_{p_n}&\in&\mathcal{N}\Big(\langle \delta v_{222}|_{p_n}\rangle,\mathrm{Var}(\delta v_{222}|_{p_n})\Big)\,.
\end{eqnarray}
The mean and variance for $\delta v_{233}$ is chosen similarly.

\paragraph{3.}  We use the symmetry conditions to determine the elements of the first two rows and columns in $\delta v_{3bc}$. The only remaining element is $\delta v_{333}$, for which we get
\begin{eqnarray}
\langle \delta v_{333}|_{p_n}\rangle&=&\Big(-v_{33}|_{p_{n}}\frac{\left\lVert\delta\phi^a\right\lVert}{\Lambda_{\text{h}}}-\langle \delta v_{331}|_{p_n}\rangle \delta\tilde{\phi}^1-\langle \delta v_{332}|_{p_n}\rangle \delta\tilde{\phi}^2-v_{33c}|_{p_n}\delta\tilde{\phi}^c\Big){\delta\tilde{\phi}^3}^{-1}  \,,\\
\mathrm{Var}(\delta v_{333}|_{p_n})&=&\Big(\frac{(1+\delta_{33})\left\lVert\delta\phi^a\right\lVert\sigma^2}{\Lambda_{\text{h}}}-(v_{33}| _{p_{n}}\frac{\left\lVert\delta\phi^a\right\lVert}{\Lambda_{\text{h}}})^2 \nonumber \\
& &
-\mathrm{Var}(\delta v_{331}|_{p_n})(\delta\tilde{\phi}^1)^2-\mathrm{Var}(\delta v_{332}|_{p_n})(\delta\tilde{\phi}^2)^2\Big){\delta\tilde{\phi}^3}^{-2} \,,\\
\delta v_{333}|_{p_n}&\in& \mathcal{N}\Big(\langle \delta v_{333}|_{p_n}\rangle,\mathrm{Var}(\delta v_{333}|_{p_n})\Big)\,.
\end{eqnarray}
The above procedure is summarized in Tab.~\ref{threeDexample} and easily implemented numerically for any $D$.

\begin{table}
\begin{center}
\begin{tabular}{ | l | c | r | }
 \hline
  $\textcolor{red}{\delta v_{111}}$ & $\delta v_{112}$ & $\delta v_{113}$ \\
  \hline
  $\delta v_{121}$ & $\textcolor{red}{\delta v_{122}}$ & $\delta v_{123}$ \\
  \hline
  $\delta v_{131}$ & $\delta v_{132}$ & $\textcolor{red}{\delta v_{133}}$ \\
   \hline
\end{tabular}\,,

\vspace{1cm} 

\begin{tabular}{ | l | c | r | }
 \hline
  $\delta v_{211}$ & $\textcolor{red}{\delta v_{212}\checkmark}$ & $\delta v_{213}$ \\
  \hline
  $\textcolor{red}{\delta v_{221}\checkmark}$ & $\textcolor{red}{\delta v_{222}}$ & $\delta v_{223}$ \\
  \hline
  $\delta v_{231}$ & $\delta v_{232}$ & $\textcolor{red}{\delta v_{233}}$ \\
   \hline
\end{tabular}\,,

\vspace{1cm}

\begin{tabular}{ | l | c | r | }
 \hline
  $\delta v_{311}$ & ${\delta v_{312}}$ & $\textcolor{red}{\delta v_{313}\checkmark}$ \\
  \hline
  $\delta v_{321}$ & $\delta v_{322}$ & $\textcolor{red}{\delta v_{323}\checkmark}$ \\
  \hline
  $\textcolor{red}{\delta v_{331}\checkmark}$ & $\textcolor{red}{\delta v_{332}\checkmark}$ & $\textcolor{red}{\delta v_{333}}$\\ 
   \hline
\end{tabular}\,.

\caption{The black entries of the tables are set to zero. The red ones have non-zero values that are drawn from distributions provided in Sec.~\ref{Sec:algorithm}. The check-mark indicates that an element is chosen according to the symmetry conditions. \label{threeDexample}}
\end{center}
\end{table}

\section{Generating Axionic Random Potentials \label{Sec:axionic}}
One possible application of DBM potentials is the modelling of globally defined potentials in cosmology/string theory by locally defined ones. Such an approach is motivated by the need for numerical experiments in the large $D$ limit. To this end, the locally defined potential needs to have identical statistical properties as the global one. In this section, we explain at the example of a simple axionic potential used in prior work \cite{Higaki:2014pja,Higaki:2014mwa,Wang:2015rel} some of the necessary steps as well as a severe problem (see \cite{Baumann:2014nda,Marsh:2015xka} for reviews on inflationary and axionic cosmology): potentials generated via DBM are not bounded, rendering them of limited use without further modification if the aim is to model a bounded potential. To be concrete, while the original DBM potentials reviewed in Sec.~\ref{DBMreview} have a bounded Hessian, they show a runway behaviour for the gradient and the potential as the traversed distance in field space increases. The reason for this known, yet hardly ever mentioned, behaviour is simple: both perform, in essence, a random walk, which causes on average a growth of the potential's and gradient's variance  ($\propto N_{\text{steps}}^3$ for the potential, as we shall see below). The same growth is present in all cases investigated in this note.

The free parameters of DBM are the vertical scale $\Lambda_{\text{v}}$, the horizontal correlation length $\Lambda_{\text{h}}$ of the Hessian and the concrete probability distribution function of the Hessian's elements (taken to be Gaussian in the original proposal expained above). The vertical scale is easily read of any globally defined potential and the potential's horizontal correlation length follows relatively straightforward as well. However, since the latter is somewhat less trivial to compute than the former and usually not identical with $\Lambda_{\text{h}}$, we provide an examble in Sec.~\ref{correlationlength} for the axionic potential investigated in \cite{Wang:2015rel}. 

\subsection{The Correlation Length \label{correlationlength}}
Consider an axionic potential defined via \cite{Wang:2015rel}
\begin{equation}
 V=\sum_{J_1,\dots,J_D=1}^{\tilde{n}} \Lambda_{J_1,\dots,J_D}\Big(1-\cos(\sum_{j=1}^D \frac{3J_j}{\tilde{n}} \phi_j+\theta_{J_1,\dots,J_D})\Big).
\label{simpledefpotential}
\end{equation}
where $D$ is the number of axions, $n=\tilde{n}^D$ is the number of shift symmetry breaking sources from non-perturbative effects, $ \Lambda_{J_1,\dots,J_D}$ sets the strength of the source terms, $\phi_j$ denote the axions with decay constants $f_j$ that are included in the mixing matrix $n_{ij}$ which are set deterministically as $n_{ij} \equiv 3J_j/\tilde{n}$ to enable comparison with analytic results in \cite{Wang:2015rel}, and $\theta_{J_1,\dots,J_D}$ are the relative phases between different source terms. 

In line with \cite{Wang:2015rel}, we let $J_j$ run from $1$ to $\tilde{n}$. Since results do not depend on the random phase factors, we set them to zero.
For the overall scale of the potential, we choose a normal distribution for the $\Lambda_{J_1,\dots,J_D}$ with mean
\begin{eqnarray}
\mu_{\Lambda}\equiv \frac{1}{\tilde{n}^D}=\frac{1}{n}
\end{eqnarray}
and standard deviation
\begin{eqnarray}
 \sigma_{\Lambda}&\equiv& \frac{a}{\tilde{n}^D}\,,\\
 a&\equiv& 0.1\,.
\end{eqnarray} 
After picking the $\Lambda_{J_1,\dots,J_D}$, we re-scale the potential such that
\begin{equation}
 \sum_{J_1,\dots,J_D=1}^{\tilde{n}} \Lambda_{J_1,\dots,J_D}= 1\,.
\end{equation}
Hence, the potential at the global minimum is $V=0$ and at the maximum $V=2$.
To define a correlation function, let us introduce the normalized variable
\begin{eqnarray}
f(\vec{\phi}) \equiv \frac{V(\vec{\phi})\, - <V(\vec{\phi})>}{<V(\vec{\phi})>}\,.
\end{eqnarray}
Since $ \, <V(\vec{\phi})>=1$, the above definition becomes $f=V(\vec{\phi})-1$. A two point correlator can be defined as
\begin{eqnarray}
C(r) \equiv \frac{<f(\vec{\phi}) \cdot f(\vec{\phi}+\vec{r})>}{<f(\vec{\phi})^2>}\,,
\end{eqnarray}
Since the potential and thus $f$ is isotropic, $C$ only depends on the distance $r$ in field space. Note that $ C\rightarrow 1$
as $r \rightarrow 0 $ and $C\rightarrow 0 $ for large $r$. It is useful to introduce the Fourier space ``power spectrum''
\begin{eqnarray}
<\tilde{f}(\vec{k})\cdot \tilde{f}(\vec{k'})>&=&\frac{A}{B}P_f(\vec{k})\delta^D(\vec{k}+\vec{k'})\,,
\label{correlatorfull}
\end{eqnarray}
so that
\begin{eqnarray}
<f(\vec{\phi})\cdot f(\vec{\phi}+\vec{r})>= B \int    P_f(\vec{k}) e^{i\vec{k}\cdot \vec{r}}d\vec{k}\,,
\label{xirelation}
\end{eqnarray}
where
\begin{eqnarray}
\tilde{f}(\vec{k})&=&\sqrt{A} \int f(\vec{\phi})e^{-i\vec{k}\cdot \vec{\phi}}d\vec{\phi}\,.
\end{eqnarray}
We leave the normalization constants $A$ and $B$ unspecified, since they cancel out in the correlator $C$ anyhow. Due to $\delta^D(\vec{k}+\vec{k'})$ in (\ref{correlatorfull}), we get
\begin{eqnarray}
<\tilde{f}(\vec{k})\cdot \tilde{f}(\vec{k'})>&=&<\tilde{f}(\vec{k})^2> \,\\
\nonumber &=&A\Big(<\sum_{J_1,\dots,J_D=1}^{\tilde{n}}\sum_{{K}_1,\dots,{K}_D=1}^{\tilde{n}}\Lambda_{J_1,\dots,J_D}\Lambda_{{K}_1,\dots,{K}_D}\prod_{j=1}^D\delta(k_j)>\,\\
\nonumber &&+\frac{1}{4}<\sum_{J_1,\dots,J_D=1}^{\tilde{n}}\Lambda_{J_1,\dots,J_D}^2\big(\prod_{j=1}^D\delta(k_j-n_{ij})+\prod_{j=1}^D\delta(k_j+n_{ij})\big)>\,\\
&&-2<\sum_{J_1,\dots,J_D=1}^{\tilde{n}}\Lambda_{J_1,\dots,J_D}\prod_{j=1}^D\delta(k_j)>+<\prod_{j=1}^D\delta(k_j)>\Big)\,,
\label{fksquare}
\end{eqnarray}
which can be simplified to
\begin{eqnarray}
<\tilde{f}(\vec{k})^2>&=&
A \frac{1+a^2}{4n}<\prod_{j=1}^D\delta(k_j-n_{ij})+\prod_{j=1}^D\delta(k_j+n_{ij})>\,.
\end{eqnarray} 
Note that $n_{ij}=3J_j/\tilde{n}$ does not depended on the index $i$. Inserting the above expression into (\ref{xirelation}) and evaluating the integral yields
\begin{eqnarray}
<f(\vec{\phi})\cdot f(\vec{\phi}+\vec{r})> \propto \Big(
\frac{1+a^2}{2n}<\cos(\sum_{j=1}^D n_{ij}\cdot r_j)>\Big)\,,
\end{eqnarray}
so that the correlator becomes
\begin{eqnarray}
C(\vec{r})=<\cos(\sum_{j=1}^D n_{ij}\cdot r_j)>\,.
\end{eqnarray}
The correlator may be expanded in a Taylor series if the correlation length is small, but it is simple to evaluate the sum numerically anyhow.
Evidently, the potentials correlation length for the axionic case under consideration is independent of number of fields $D$ and the number of shift symmetry breaking terms $\tilde{n}$. Evaluating the correlation length numerically for the above mentioned distributions of parameters yields the ensemble average  $\Lambda_{\text{h}}(V)\sim 0.75$.

However, $\Lambda_{\text{h}}(V)$ is in general not identical with the correlation length of the Hessian, $\Lambda_{\text{h}}(V_{ab})$. One may either adjust the latter iteratively in a concrete DBM potential to match the desired $\Lambda_{\text{h}}(V)$ of the globally defined potential or compute $\Lambda_{\text{h}}(V_{ab})$ semi-analytically, as we did with $\Lambda_{\text{h}}(V)$ above, to match it directly. It should be noted that it is usually not possible to match  the potential's and the Hessian's correlation length simultaneously, since only one free parameter can be adjusted.

\subsection{DBM Potentials with Adjusted Mean of the Hessian's Entries}
The entries of the Hessian for the axionic potential in (\ref{simpledefpotential}) are, to a good approximation, Gaussian random variable with mean
\begin{eqnarray}
\mu_{kl} \approx \dfrac{9}{\tilde{n}^2}(1-V_{\text{c}}) \label{meanHessianentries}
\end{eqnarray}
 and standard deviation
\begin{equation}
 \sigma_{kl}  = \begin{cases} \sigma_{\text{dia}} \approx \dfrac{3.98}{\sqrt{n}} &\mbox{for $k=l$,} \\
\sigma_{\text{offdia}} \approx  \dfrac{2.2}{\sqrt{n}} &\mbox{for $k \neq l$,}
\end{cases} 
\label{sigmadiaandoff}
\end{equation}
see \cite{Wang:2015rel} for details. Here, $V_c$ is the value of the potential at the point in field space at which the Hessian is to be evaluated. Note that the term $1-V_c$ makes maxima more likely for $V>1$ and minima for $V<1$.
So, instead of the condition in (\ref{condition1})
\begin{eqnarray}
\langle \delta  v_{ab}|_{p_n}\rangle =-v_{ab}|_{p_{n-1}}\frac{\left\lVert\delta\phi^a\right\lVert}{\Lambda_{\text{h}}}
\end{eqnarray}
we need to demand
\begin{eqnarray}
\langle \delta v_{ab}|_{p_n} \rangle =\Big(\frac{9}{\tilde{n}^2}(1-v_0|_{p_{n-1}})-v_{ab}|_{p_{n-1}}\Big)\frac{\left\lVert\delta\phi^a\right\lVert}{\Lambda_{\text{h}}}\,.
\label{axionconstrainc1}
\end{eqnarray}
Enforcing (\ref{axionconstrainc1})  guarantees that the entries of the Hessian acquire the mean in (\ref{meanHessianentries}). We do not impose the different standard deviations in (\ref{sigmadiaandoff}).

\begin{figure}[tp]
\centering
\includegraphics[width=0.75\textwidth]{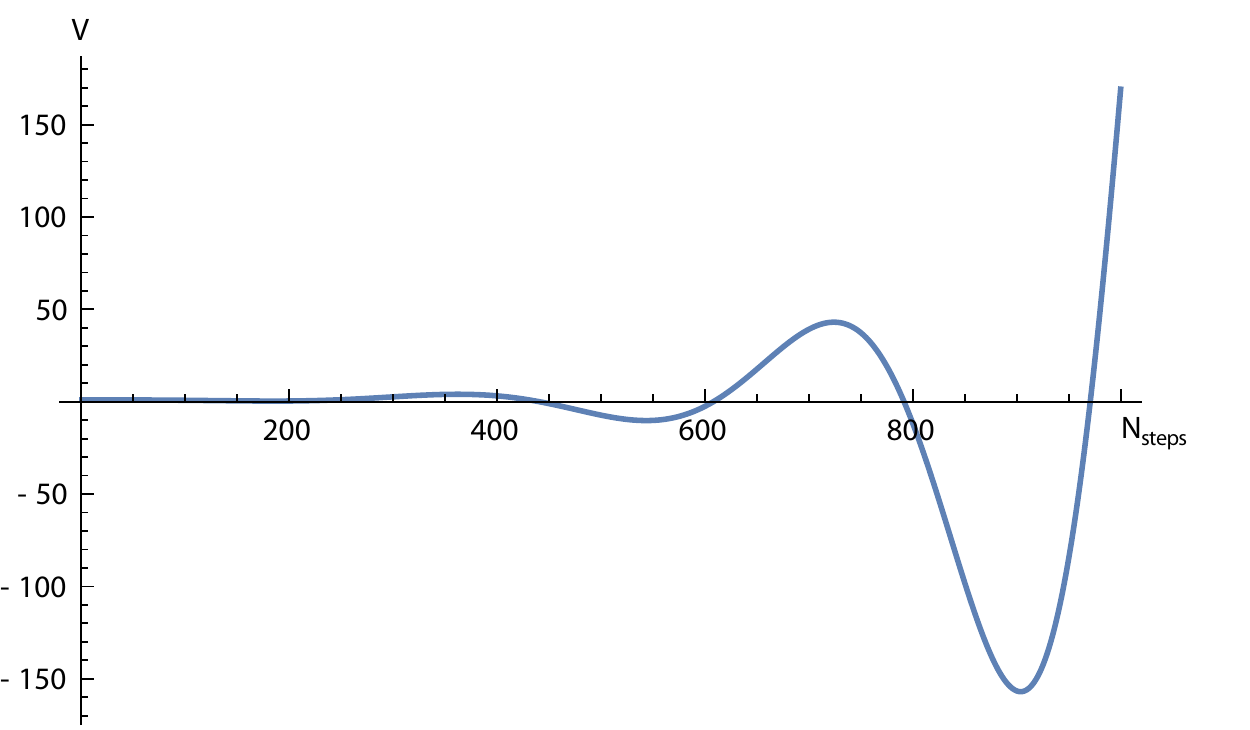}
\caption{Locally defined DBM potential with a mean of the Hessian's entries in  (\ref{meanHessianentries}), $n=9$ and $D=3$ over $1000$ steps. Each step corresponds to $1\%$ of the Hessians correlation length. The corresponding global potential is bounded between $0$ and $2$, while the locally constructed one shows oscillations with increasing amplitude.}
\label{fig:c1axionnomodi}
\end{figure}

Random potentials generated this way show a common feature, namely, they oscillate with a growing amplitude, as shown in Fig.~\ref{fig:c1axionnomodi}. After a few correlation lengths, the potential grows to an unreasonable height, far beyond the bounds of the globally defined potential. 

Let's see if we can pinpoint the origin of this feature: Our starting point for DBM is a small $v_{ab}|_{p_0}$ with $\left\lVert\delta\phi^a\right\lVert/\Lambda_{\text{h}}=0.01$ and $\Lambda_{\text{v}}^4\sqrt{D}=1$. Thus, the first term $9(1-v_0)/\tilde{n}^2$ contributes much more than the second term, $-v_{ab}$. The second term tries to compensate the effect of the last step while the first term changes it´s sign (and thus whether to push the potential upwards or downwards\footnote{For the sake of simplicity, we let the potential evolve in one fix positive direction in this argument, so the sign of $v_{ab}|_{p_n}$ translates directly into the sign of $\delta v_0|_{p_{n+1}}$.}) after crossing $v_0=1$.
Since the second term is smaller the first one, the DBM algorithm will always tend to push the potential downward(upwards), if $v_0$ is above(below) one. Thus, if we start with $v_0>1$, a large gradient results at $v_0=1$. After crossing $1$ from above, the DBM algorithm tries to push the potential upwards, but the large gradient at $V=1$ needs to be compensated in order to turn around, which in turn takes a longer distance the higher the magnitude of the gradient is. Therefore, once the potential reaches the next minimum, it is at a lower value than before which causes an even larger gradient at the next crossing of $v_0=1$. This mechanism leads to the observed oscillations with a growing amplitude. 

\begin{figure}[tp]
\centering
\includegraphics[width=0.75\textwidth]{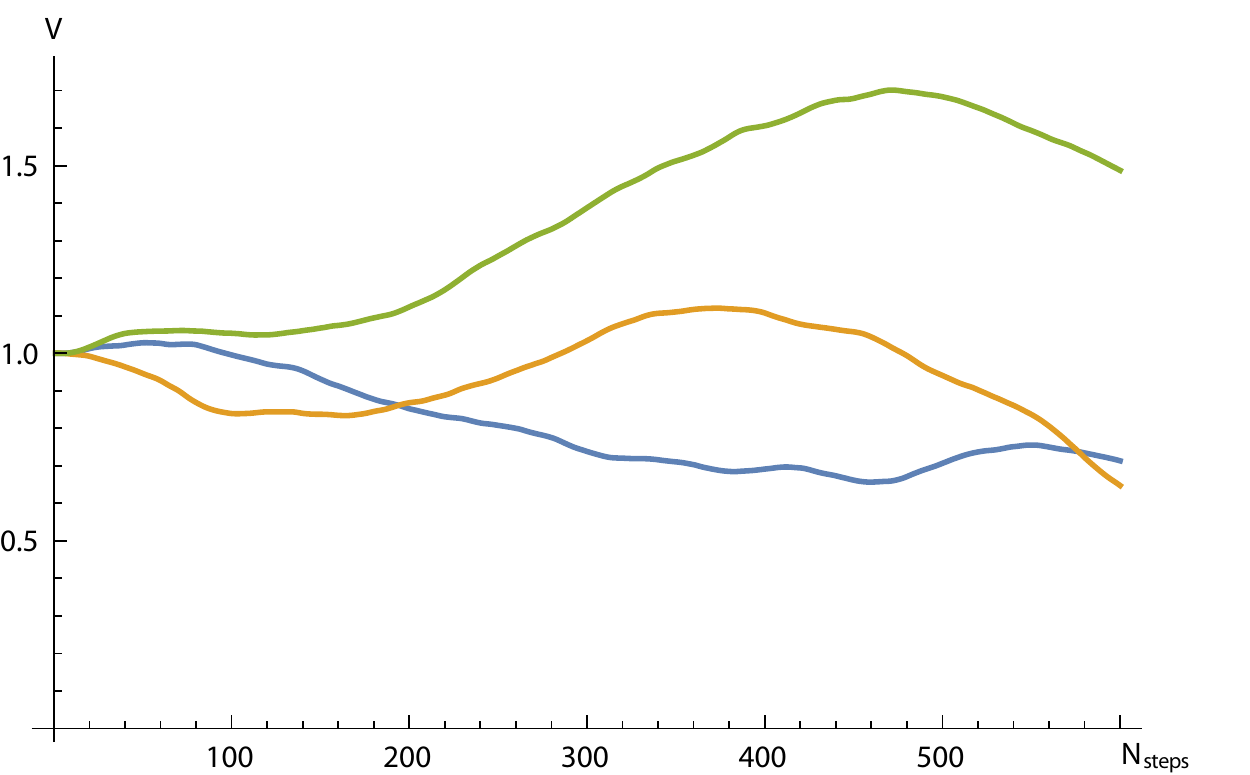}
\caption{Three exemplary realisations mimicing a $\tilde{n}=5$ and $D=3$ axionic potential over $600$ steps, each of which corresponding to $1\%$ of the Hessian's correlation length. The extended DBM potential is generated with the mean and variance in (\ref{modimean}) and (\ref{modivar}) with $A=36,B=10,C=3$. Note the still present unstable growth of the potential's variance  $\propto N_{\text{steps}}^3$ that can't be removed by this simple extension of DBM. }
\label{fig:v0example}
\end{figure}

The growth of $V$ can be quantified by the variance of the potential as a function of $N_{\text{steps}}$. For example, a numerical study to mimic the axionic landscape with $D=3, \tilde{n}=5$ via a DBM yields
\begin{eqnarray}
 \text{Var}(V)\approx-130 + 1.1 \times 10^{-7} N_{\text{steps}}^3\,.
 \label{dispersion}
\end{eqnarray}
The dependence $\propto N_{\text{steps}}^3$ is generic.

Thus, such DBM potentials do a poor job at modeling the axionic landscape: while the Hessian has the proper distribution, the potential and the gradient are completely dominated by these unstable oscillations. The same artifact is present for the Extended Dyson Brownian Motion algorithm of Sec.~\ref{Sec:extDBM}. Thus, such potentials can not be used to draw any conclusion with respect to the distribution of extrema or the likelihood of achieving cosmological inflation on axionic landscapes.

It is possible to modify the DBM-algorithm to keep the potential under control for a while by increasing the magnitude of the second term and thus making the instantaneous back reaction stronger. However, the unstable growth of the potential's variance $\propto N_{\text{steps}}^3$ is still present, as in all DBM potentials considered in the literature, see the next section. 

\subsection{Modifying DBM,  a Case Study \label{axionmodi}}

Let us consider the ad hoc introduction of a free parameter into the mean of $\delta v_{ab}$, while keeping the variance unchanged,
\begin{eqnarray}
\langle \delta  v_{ab}|_{p_n}\rangle&=&-A*v_{ab}|_{p_{n-1}}\frac{\left\lVert\delta\phi^a\right\lVert}{\Lambda_{\text{h}}}\,,\label{modimeanori}\\
\mathrm{Var}(\delta v_{ab}|_{p_n})&=&\frac{(1+\delta_{ab})\left\lVert\delta\phi^a\right\lVert\sigma^2}{\Lambda_{\text{h}}}-(v_{ab}|_{p_{n-1}}\frac{\left\lVert\delta\phi^a\right\lVert}{\Lambda_{\text{h}}})^2\,.\label{modivarori}
\end{eqnarray}
For appropriately chosen values of $A$, it is usually possible to reduce the numerical pre-factor in the growth of the variance considerably, but $\sigma(v_{ab})$ is altered as well; furthermore, the dependency on  $\propto N_{\text{steps}}^3$ remains. To illustrate this point, let us try to mimic an axionic landscape with $D=3,\tilde{n}=5$.

\begin{figure}[tp]
\centering
\includegraphics[width=0.75\textwidth]{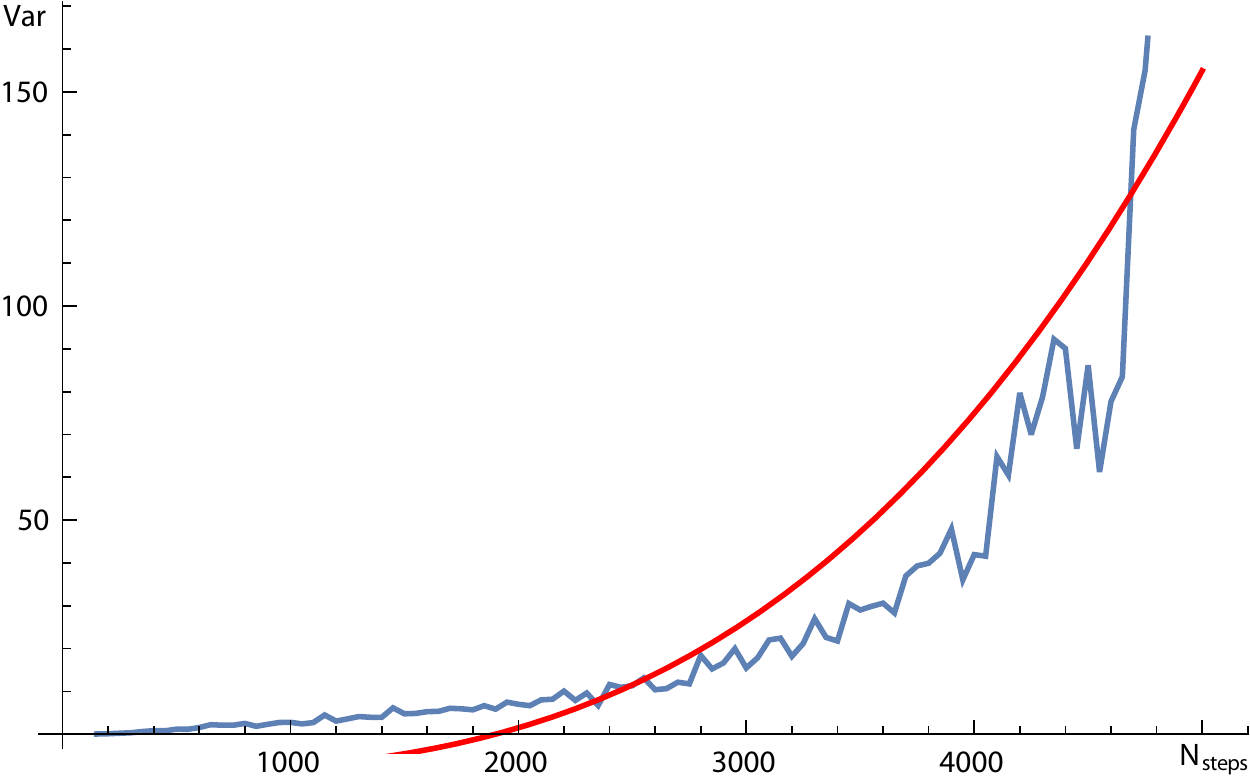}
\caption{$\text{Var}(V)$ over the number of steps for extended DBM potentials with mean and variance in (\ref{modimean}) and (\ref{modivar}) respectively, as well as $A=36,B=10$ and $C=3$. Red line is the fit in (\ref{fit1}) and the blue line the numerical result. Note the unstable growth of the potential's variance  $\propto N_{\text{steps}}^3$. }
\label{fig:dispersionaxionmodified}
\end{figure}

In line with (\ref{axionconstrainc1}), we introduce the free parameter $A$ into the mean of $\delta v_{ab}$,
\begin{eqnarray}
\langle \delta v_{ab}|_{p_n}
\rangle =\Big(\frac{9}{\tilde{n}^2}(1-v_0|_{p_{n-1}})-A*v_{ab}|_{p_{n-1}}\Big)\frac{\left\lVert\delta\phi^a\right\lVert}{\Lambda_{\text{h}}}\,.
\label{modimean}
\end{eqnarray}
To compensate for the alteration of the Hessians variance, we allow for two additional free parameters, $B$ and $C$,
\begin{eqnarray}
\mathrm{Var}(\delta v_{ab}|_{p_n})&=&B\times \Big(\frac{C(1+\delta_{ab})\left\lVert\delta\phi^a\right\lVert \sigma^2}
{\Lambda_{\text{h}}}-(v_{ab}|_{p_{n-1}}\frac{\left\lVert\delta\phi^a\right\lVert}{\Lambda_{\text{h}}})^2\Big)\,,
\label{modivar}
\end{eqnarray}
We first optimize $A$ to match the desired standard deviation of the potential for a longer stretch in field space: starting from an initial value $A=20$ we increase $A$ in steps of one and generate $60$ realizations for each case. All realizations run for $600$ steps. We demand that $v_0$ lies in the interval $[0,2]$. Once $v_0$ exceeds this range, we delete this realization and start a new run. We follow with the computation  of the potential's standard deviation for the 600 steps, and calculate the mean value of the 60 realizations. We stop searching once a suitable value of $A$ is found, for which we demand that the averaged standard deviation differs from the potential's desired one (see \cite{Wang:2015rel}) no more than $0.01$, i.e.~about $ 50 \%$ relative deviation. The smallest value of $A$ found via this algorithm is $A=36$ in our particular case.

\begin{figure}[tp]
\centering
\includegraphics[width=0.75\textwidth]{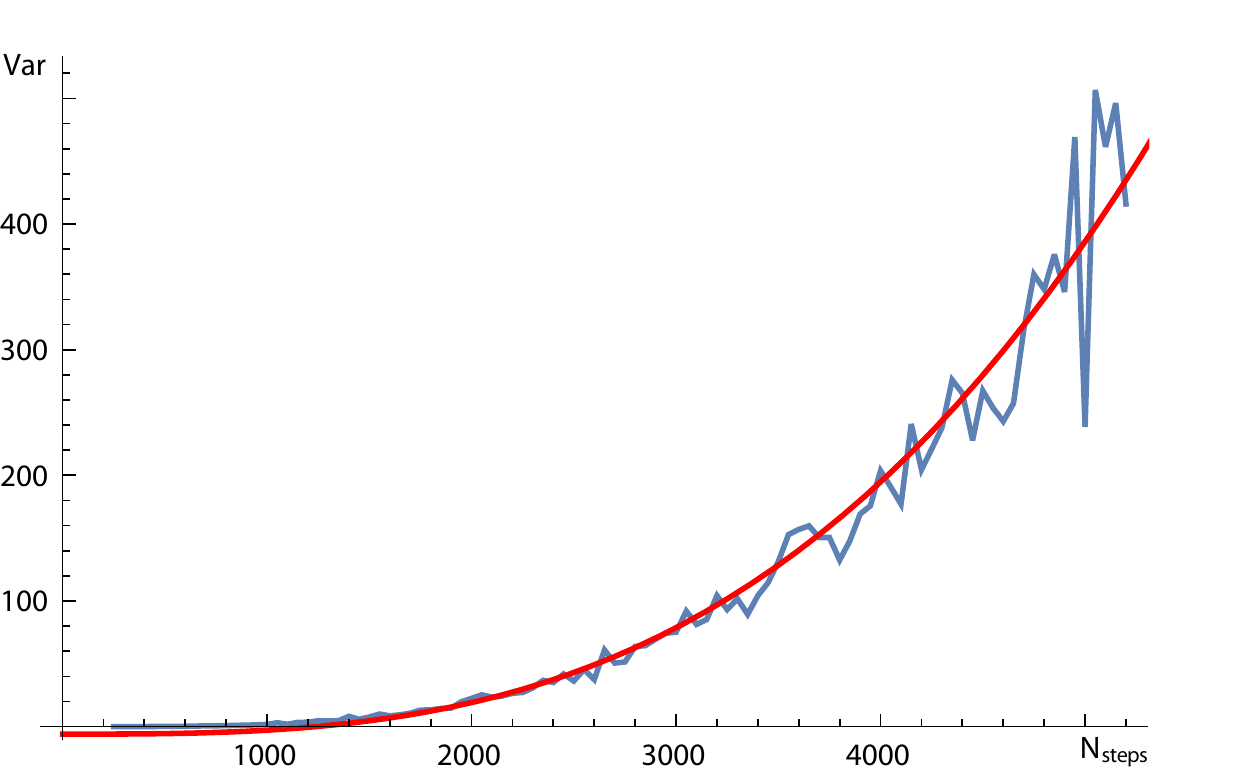}
\caption{$\text{Var}(V)$ over the number of steps for the original DBM potentials with mean and variance in (\ref{modimeanori}) and (\ref{modivarori}) respectively  as well as $A=36,B=1$ and $C=1$. Red line is the fit in (\ref{fit2}) and the blue line the numerical result. Note the unstable growth of the potential's variance  $\propto N_{\text{steps}}^3$.}
\label{fig:dispersionmodified}
\end{figure}

Holding $A$ fixed, we vary $B$ and $C$ at the same time. We start with $B=3, C=2$ and compute both, $\sigma(V)$ and $\sigma(v_{ab})$. For their deviations from the desired values we demand the difference to be smaller than $0.1$ for  $\sigma(v_{ab})$ and  smaller than $0.05$ for $\sigma(V)$. For a fixed $B$, we increase $C$ in steps of one. If $C$ grows over $5$ and the two conditions haven't been satisfied, we increase $B$ by one and continue with the same procedure. The set of parameter we found via this algorithm is
\begin{eqnarray}
A=36\quad,\quad B=10\quad,\quad C=3\,.
\end{eqnarray} 
The above algorithm can of course be optimized, but we are primarily interested to provide an instructive example for which our method is sufficient.

Three exemplary realizations are shown in Fig.~\ref{fig:v0example}. Evidently, the oscillations with increasing amplitude are not present any more. However, due to the still present unstable growth of the potentials variance with $N_{\text{steps}}^3$, one can not mimic the desired landscape for large step numbers, even with this modification. Furthermore, for low step numbers, the potential is dominated by the chosen initial conditions.

To be concrete, the variance can be fitted well by
\begin{eqnarray}
\text{Var}(V)\approx -9 + 1.3 \times 10^{-9} N_{\text{steps}}^3\,, \label{fit1}
\end{eqnarray}
see Fig.~\ref{fig:dispersionaxionmodified}, showing clearly the dependence $\propto N_{\text{steps}}^3$.
 
Performing the same modification on the original DBM with mean and variance in (\ref{modimean}) and (\ref{modivarori}) respectively, that is $A=36$, $B=1$ and $C=1$, shows the same qualitative behaviour: while the prefactor is reduced by  the choice of $A$, the proportionality with $\propto N_{\text{steps}}^3$ remains. A fit leads to
\begin{eqnarray}
\text{Var}(V)\approx-6 + 3 \times 10^{-9}  N_{\text{steps}}^3\,, \label{fit2}
\end{eqnarray}
see Fig.~\ref{fig:dispersionmodified}. The same unstable growth can be observed in $C^2$-potentials generated via the algorithim introduced in Sec.~\ref{sec:C2potentials}. Similarly, the variance of the gradient is growing with the traversed distance in field space. Thus, due to this unstable growth, potentials generated via DBM have limited use if the goal is to mimic a globally defined one for an extended stretch in field space. For a local study, for example to study effects of a mild random component to the potential near a saddle point on which inflation takes place, DBM potentials can be used, see e.g. \cite{Dias:2016slx}. One needs to be cautious though, since the presence of a flat region suitable for inflation is due to ones initial conditions. Also, due to the growth of the gradient's variance, we expect that the end of inflation is strongly affected by the method by which the potential is generated, i.e. Dyson Brownian Motion.

\section{Conclusions}

We provided a computationally efficient extension of the Dyson Brownian Motion algorithm to generate random function in $C^2$ locally, as desirable for certain applications in cosmology. We further showed at the example of a simple globally defined potential that DBM potentials fail to recover basic features of the globally defined ones, due to the presence of an unstable growth of the gradient's and the potential's variance. We also showed that a minor ad hoc modification of the algorithm can weaken the unstable behaviour, such that a DBM potential can mimic a globally defined one reasonable well for some time. However, the unstable growth is still present. 

Thus, in their current form, DBM potentials can be used to model landscapes in String Theory or to draw conclusions in cosmological settings only for regions not much bigger than their coherence length. 

\acknowledgments

We would like to thank D.~Marsch, N.~M.~Nguyen, F.~G.~Pedro and A.~Westphal for discussions.

-----------------------------------------------------------------

\end{document}